\documentclass[journal]{IEEEtran}
\usepackage{cite}
\usepackage{amsmath,amssymb,amsfonts,mathtools}
\usepackage{algorithm}
\usepackage{algpseudocode}
\usepackage{graphicx}
\usepackage{textcomp}
\usepackage{xcolor}
\usepackage{multicol}
\usepackage{wrapfig}
\usepackage{float}
\usepackage{stfloats}
\usepackage[caption=false,font=normalsize,labelfont=sf,textfont=sf]{subfig}

\begin{document}

\title{Adaptive Beam Broadening for DoA Estimation in Dynamic and Resource-Constrained DFRC Systems}

\author{\IEEEauthorblockN{D.R. Raghavendra, V.V. Reddy, and A. Mishra}\\
\IEEEauthorblockA{International Institute of Information Technology, Bangalore, India.\\
Email: \{raghavendra.dr, vinod.reddy, amrita.mishra\}@iiitb.ac.in\thanks{A. Mishra would like to acknowledge the financial support of Infosys Foundation Career Development Chair Professorship grant.}}}

\maketitle

\begin{abstract}

Dual-function radar communication (DFRC) systems incorporate both radar and communication functions by sharing spectrum, hardware and radio frequency (RF) chains. In this work, we consider a conceptual DFRC scheduler model which shares RF chains between radar and communication functions. If such a scheduler is tuned for prioritizing communication performance, the RF chains and time allocated to radar are less and varying. We propose a practical, low-latency and resource-aware technique for sensing the entire field-of-view (FOV) and Direction-of-Arrival (DoA) estimation in such settings by leveraging time-sliced beam allocation along with adaptive windowing. This results in a balanced cumulative array factor over the FOV thereby ensuring better DoA estimation reliability. Extensive simulation studies show that the technique has consistent target detection and angle estimation performance in all directions and adapts to varying resource availability with time.
\end{abstract}

\IEEEpeerreviewmaketitle

\section{Introduction}

Future applications such as autonomous driving, unmanned aerial vehicles (UAVs) and others require reliable, low-latency sensing and communication. Traditionally, sensing and communication have developed along different trajectories and a paradigm shift towards a unified solution has been of interest. This is accomplished by DFRC systems in which sensing and communication coexist \cite{zheng2019dfrc} \cite{Hassanien2019DualFunctionRC}. Coexistence involves the sharing of resources like spectrum, hardware systems and RF signal processing chains \cite{Liu2020dfrc} \cite{Liu2017TowardDR} \cite{Sturm2011waveform}. From hardware systems perspective, resources such as RF chains and antennas need to be shared.
\\

Due to higher bandwidth translating to higher data rates and low latency, mmWave phased array is seen as a key enabler for achieving the goals of DFRC in 5G/6G. Silicon-based mmWave phased arrays offers the advantages of small size, low cost and enhanced integration capabilities on a single chip. Recently, important advancements such as increased beam count, integrated beam memory arrays and precise beam steering with both phase and amplitude control have been developed and practically demonstrated \cite{Sadhu2017}. As a result, these systems are expected to be widely incorporated in 5G base stations (BSs) and also drive the future complex and adaptive multi-function systems in 6G.
\\

In 5G cellular systems, the BS shares the radio resources by time division into slots encompassing Synchronization Signal Blocks (SSBs), Initial
Access (IA) phases and data transmission phases for user equipments (UEs). These processes are orchestrated by a centralized scheduler responsible for resource allocation, timing coordination, and interference management. In 6G DFRC systems, the scheduler is expected to assume an even more prominent and context-aware role to manage increasingly dynamic, heterogeneous, and latency-sensitive network environments. Assuming communication with UEs takes precedence over sensing, the number of required RF chains for communication varies from frame to frame. We then pose the question - when communication takes precedence thereby \textit{limiting the RF chains} available for sensing, how can sensing across full FOV be achieved in a \textit{single sensing frame}?
\\

The most common approach employed for sensing the entire FOV is sequential sensing, wherein different regions are sensed with the available beams in a sequential fashion until the entire FOV is covered \cite{belfiori2012random} \cite{Billam1997Time}. Other approaches include increasing the beamwidth of each beam by using sub-arrays \cite{GrahamSubArray2013} \cite{RajSubArrays2012}. Alternatively, analog beamforming weights are completely redesigned using array pattern synthesis \cite{gershman1999robust} \cite{Nai2010BeamPatternSynthesis}. The above techniques have been typically designed for radar-only or communication-only systems. 
\\

We consider time and resource-constrained shared DFRC setting and propose sensing the scene in such scenarios using a multi-stage adaptive windowing technique. This technique has the desirable property that it ensures reliable sensing in low-resource scenarios and in the best case scenario, when enough resources are available, its behavior approaches that of sequential sensing. It is practical to implement given the recent advances of beamformer ICs such as real-time phase and amplitude control, beam tables and integrates smoothly into the RF signal processing chain.
\section{System and Signal model}
\subsection{System Model}
\begin{figure}[htbp]
\centering
\includegraphics[width=0.5\textwidth]{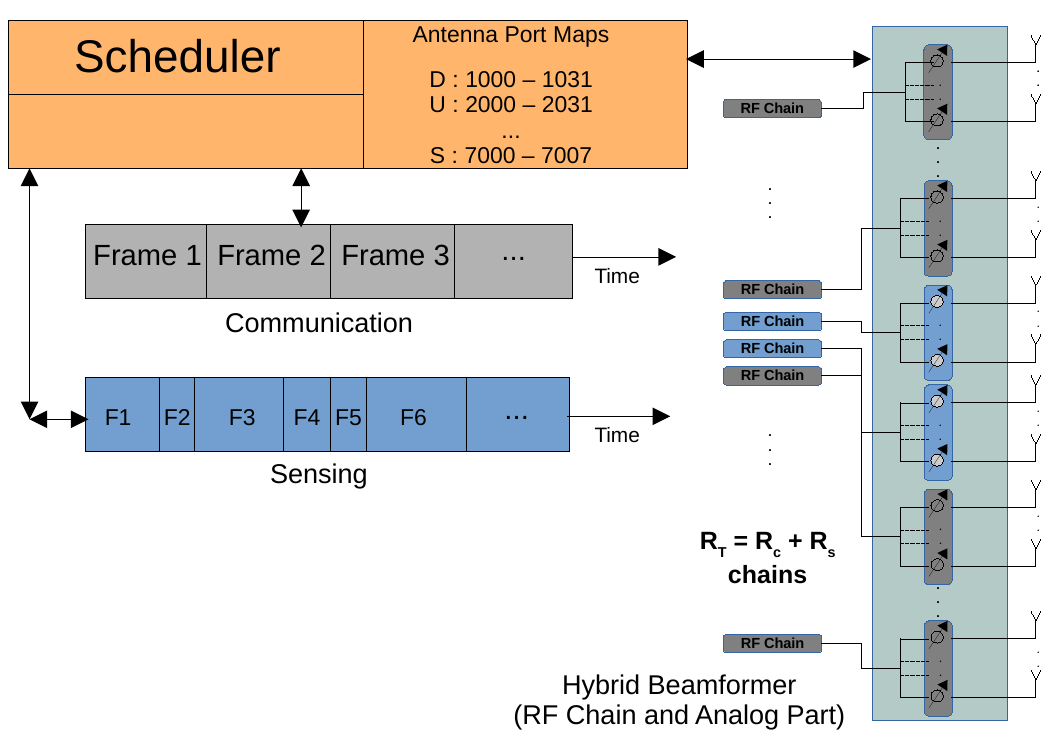}
\caption{\small{\label{fig:5G_6G_Scheduler_RF_chains} An abstract scheduler which manages communication and sensing functions by sharing RF chains. }}
\end{figure}
At the physical level, the DFRC BS hardware can be considered to have $R_T$ RF chains realized by a sub-connected hybrid beamforming architecture as shown in Fig.~\ref{fig:5G_6G_Scheduler_RF_chains}. Each RF chain feeds the signal to $L$ antenna elements through a programmable analog beamformer. Thus, each RF chain is steerable in a specific direction with desired beamwidth by appropriate analog beamformer weights. For sensing, without loss of generality, a uniform linear array (ULA) with $P$ elements is considered. From a sensing perspective, the BS is effectively a monostatic phased-array radar with varying number of transmit RF chains. 
\\

At the logical system level, we abstract out two features of a scheduler for a DFRC system 1) The scheduler dynamically manages the allocation of RF chains between radar and communication and 2) The scheduler will schedule the radar function at appropriate time instants for a dynamic number of sensing slots (F1, F2, F3,...) as shown in Fig.~\ref{fig:5G_6G_Scheduler_RF_chains}. At the start of the $i^{th}$ sensing frame, the scheduler first determines the need for $R_{C_i}$ RF chains amongst the available $R_T$ RF chains so that the communication spectral efficiency is maintained. The remaining $R_{S_i} = R_T-R_{C_i}$ RF chains are available for sensing over a sensing interval of $T_{\mathrm{S_i}}~$s as determined by the scheduler. An instance of the RF chain configuration for the $i^{th}$ sensing frame is colour coded in Fig.~\ref{fig:5G_6G_Scheduler_RF_chains}. In this instance, only two RF chains are allocated to sensing and the rest to communication.
\subsection{Signal Model}
For the $m^{th}$ RF chain connected to $L$ antennas forming a beam along $\theta_m$, the complex beamforming weights are given by
{\footnotesize
\begin{equation}\label{eq:beamformer_weights}
\mathbf{b}(\theta_m) = \Big[1, e^{-j2\pi\frac{d sin \theta_m}{\lambda}}, e^{-j2\pi\frac{d 2sin \theta_m}{\lambda}}, ..., e^{-j2\pi\frac{d (L-1)sin \theta_m}{\lambda}}\Big]^T,
\end{equation}
}
where \( \lambda  \) is the wavelength and \(d\) is the antenna spacing.
\\

Let \(M\) RF chains be allocated to sensing for a particular frame instance $i$, i.e, $R_{s_i}=M$. To span the desired FOV, $[\theta_{start}, \theta_{end}]$, the $M$ beams have to be positioned such that the $m^{th}$ beam directions are given by
\begin{equation}\label{eq:mth_beam_position}
\theta_m = \arcsin\left( \sin(\theta_{start}) + m \frac{\left(\sin(\theta_{end}) - \sin(\theta_{start})\right)}{M} \right)
\end{equation}
for $1 \leq m \leq M$. The signal transmitted by the antenna elements of this RF chain is given by 
\begin{equation}\label{eq:2}
\mathbf{x}_m(t) = \mathbf{b}(\theta_m) s_m(t),
\end{equation}
where $s_m(t)$ is the transmit signal fed to the beamformer. 
\\

Stacking the signals transmitted by all the $M$ sensing RF chains, we obtain
\begin{equation} \label{eq:tx_signal_no_window}
\mathbf{x}(t) = \mathbf{B} \mathbf{S}(t) \in \mathbb{C}^{LM \times 1}, 
\end{equation}
where \(\mathbf{B}\) is the beamforming matrix given by
\begin{equation} \label{eq:4}
\mathbf{B} = \mathrm{diag}\left([\mathbf{b}^T(\theta_1),\ldots, \mathbf{b}^T(\theta_M)]^T\right)
\end{equation}
and 
\begin{equation}
\mathbf{S}(t) = \big(\mathbf{s}(t) \otimes \mathbf{1}_{L}\big ) \in \mathbb{C}^{ML\times 1}.
\end{equation}
Here \(\mathbf{s}(t)=[s_1(t), s_2(t), \ldots, s_M(t)]^T\), \(\mathbf{1}_{L}\) is a \(L\times 1\) vector of ones and \(\otimes\) denotes the standard Kronecker product.
\\

Since angle estimation cannot be effectively performed after analog beamforming at the receive, a separate phased-array receiver is considered for sensing. The signal received by the phased-array receiver after interaction with the $k$th far-field target is given by
\begin{equation}
\mathbf{z}(t) = \sum_{k=1}^{K} \alpha_k \mathbf{a_{R}}(\phi_k) \mathbf{a_{T}}^H(\phi_k)\mathbf{x}(t-\tau_k),
\end{equation}
where \(\alpha_k\) is the strength of the \(k^{th}\) target response, \(\mathbf{a_{R}}(\phi_k)  \in \mathbb{C}^{P\times 1}\) and \(\mathbf{a_{T}}(\phi_k)  \in \mathbb{C}^{ML\times 1} \) are the steering vectors for the \(k^{th}\) target at an angle \( \phi_k \) as seen from the radar receive and transmit arrays respectively and \( \tau_k \) is the time delay associated with \(k^{th}\) target.

\subsection{Problem Formulation}

\begin{figure}[htbp]
\centering
\includegraphics[width=0.5\textwidth]{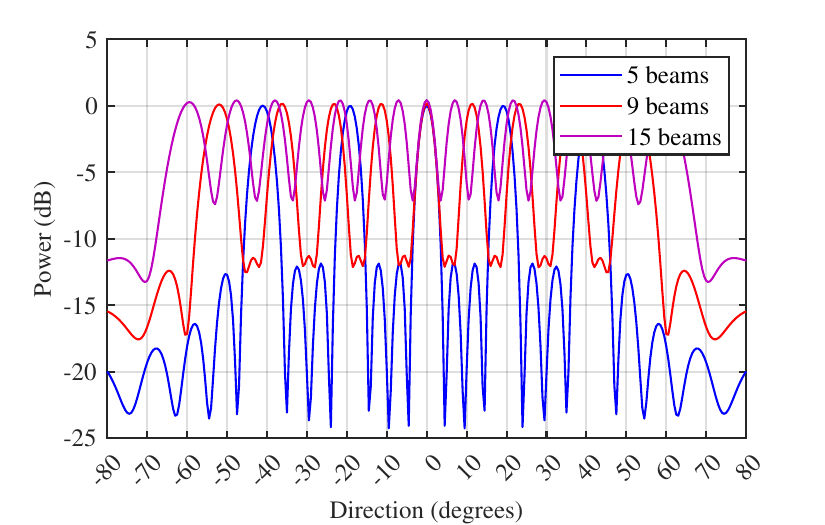}
\caption{\small{\label{fig:25_sensors_beam_patterns} Normalized beam pattern $\mathbf{f}(\theta)$ over a $[-80^{\circ}, +80^{\circ}]$ FOV for $15, 9 \text{ and } 5$ beams. Beam nulls can be observed in the beam pattern for $5$ beams.}}
\end{figure}

For a single beamformer with weights as given in (\ref{eq:beamformer_weights}), we recall that the theoretical array factor at an angle $\theta_0$ is given by
\begin{equation}\label{eq:af_theta_0}
   \mathbf{f}_{\theta_0}(\theta) = \sum_{k=1}^{L} e^{-j2\pi\frac{d}{\lambda}k(\sin{\theta} - \sin{\theta_0)}}
\end{equation}
\\
and for $R_{S_i}$ beams with beam directions chosen according to (\ref{eq:mth_beam_position}), the cumulative array factor is given by
\begin{equation}
   \mathbf{f}(\theta) = \sum_{i=1}^{R_{S_i}} \sum_{k=1}^{L} e^{-j2\pi\frac{d}{\lambda}k(\sin{\theta} - \sin{\theta_i)}}.
\end{equation}
\\
Since $R_{S_i}$ changes dynamically from frame to frame, the normalized overall beam patterns $\mathbf{f}(\theta)$ for cases with $R_{S_i} = 5$, $9$ and $15$ beams are plotted in Fig. \ref{fig:25_sensors_beam_patterns}. 
When $R_{S_i} = 5$, the effective beam pattern is observed to have power drop of nearly $25$ dB in many directions, resulting in beam null positions within the FOV. A target in one such beam nulls will go undetected unless it gets very close in range. On the contrary, the directions with beam pattern above the half-power beamwidth (HPBW), i.e., $>-3~$ dB level are suitable for target detection. Such a radar with inconsistent performance in different directions within the FOV is undesirable. A radar with comparatively lesser range but complete FOV coverage can be considered more reliable.

When $R_{S_i}= 9$, the cumulative beam pattern contains more angular region above HPBW and a raised beam pattern floor. Although better than having $R_{S_i}= 5$ beams, the inconsistency in radar detection performance still exists. 

With the coverage of FOV being of paramount importance for sensing, the problem at hand is to broaden the beams considering the available RF chains in the current sensing frame and sensing the desired FOV within $T_{S_i}~$s. In the next section, we propose a dynamic mechanism for transmit beam broadening that assures consistent performance over the entire FOV factoring in the above considerations.

\section{Proposed Dynamic Multi-Stage Adaptive Transmit Beam Broadening}

\subsection{Multi-Stage Adaptive Window Design}

At the start of the $i^{th}$ sensing frame as shown in Fig. \ref{fig:5G_6G_Scheduler_RF_chains}, the scheduler determines the duration  $\left(T_{S_i}\right)$s for which the remaining $R_{S_i}$ or $M_i$ RF chains can be allocated for sensing.

Taking into account $T_{S_i}$ and the radar's coherent processing interval $T_{CPI}$, we identify the number of sensing CPIs (stages) that can be guaranteed to complete within $T_{S_i}$ s. This enables sensing with more than $M_i$ beams across stages. Therefore, the number of stages for the $i^{th}$ frame is given by
\begin{equation} \label{eq:N_stg}
N_{stg_i} = \left \lfloor\frac{T_{S_i}}{T_{CPI}} \right \rfloor.
\end{equation}

Considering an arbitrary frame, we can drop the subscript $i$ and denote the number of stages as $N_{stg}$ and number of sensing RF chains as $M$. Then, a total of $MN_{stg}$ beams are available for sensing the environment that can be staggered over $N_{stg}$ stages within $T_{S_i}$ using the $M$ available beams. 
\\

Even with this multi-stage sensing with $MN_{stg}$ beams, beam nulls can occur in some directions depending on $M$ and $N_{stg}$. Parametrized windows are employed as adaptive filters on the analog beamformer to overcome possible beam nulls and balance power over the entire FOV. 
\\

The positions of the $m \in [1,MN_{stg}]$ beams can be obtained apriori at the onset of the frame by rewriting (\ref{eq:mth_beam_position}) as
\begin{equation}\label{eq:mnstgth_beam_position}
\theta_m = \arcsin\left( \sin(\theta_{start}) + m \frac{\left(\sin(\theta_{end}) - \sin(\theta_{start})\right)}{MN_{stg}} \right).
\end{equation}

The half-power beam width (HPBW), $\Delta\theta^{\text{hpbw}}_{m}$, for an RF chain with ULA configuration is given by \cite[pp.~47--49]{Van2002optimum}
\begin{equation} \label{eqn:hpbw_beamwidth_approx}
    \Delta\theta^{\text{hpbw}}_{m} \approx \frac{0.886 \lambda}{Ld\cos{\theta_{m}}}
\end{equation}
for a beam at $\theta_m$ and the first-null beam width (FNBW) is approximated as $\Delta\theta^{\text{fnbw}}_{m} \approx 2\Delta\theta^{\text{hpbw}}_{m}$. 
\\

The angular separation between HPBW points of the $m^{th}$ and $(m+1)^{th}$ beam is then given by
\begin{equation} \label{eqn:angular_sep_beams}
\Delta \theta^{sep}_{m} = \left(\theta_{m+1} - \frac{\Delta\theta^{\text{hpbw}}_{m+1}}{2}\right) - \left(\theta_{m} + \frac{\Delta\theta^{\text{hpbw}}_{m}}{2}\right).
\end{equation}
\begin{figure}[htbp]
\centering
\includegraphics[width=0.5\textwidth]{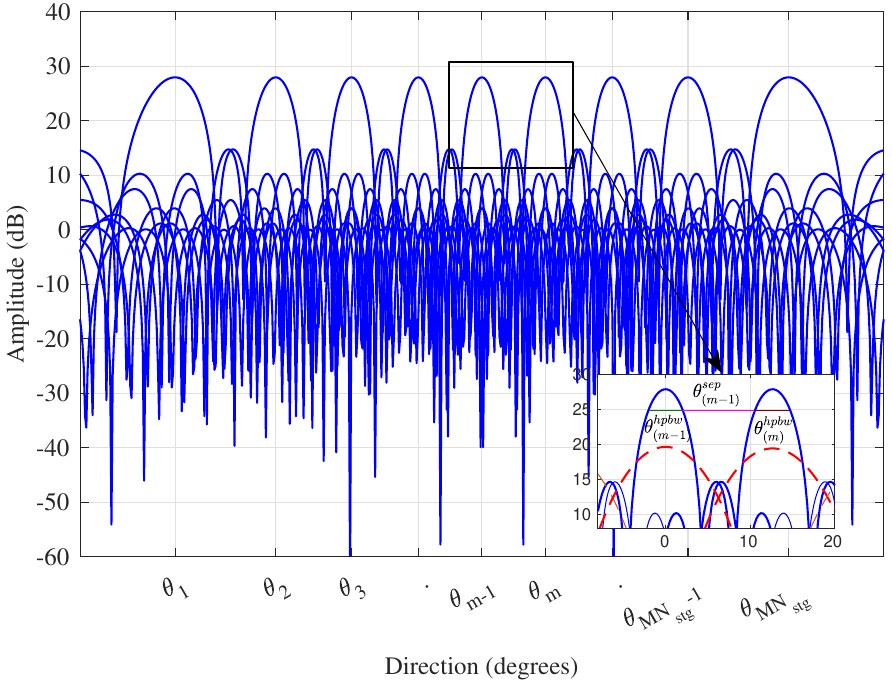}
\caption{\small{\label{fig:beams_hpbw} Beam pattern for $M=3$ and $N_{stg}=3$, i.e, $MN_{stg} =9$ beams. Inset shows the HPBWs and HPBW separation between $m^{th}$ and $(m-1)^{th}$ beams along with hypothetical broadened beams. }}
\end{figure}
If $\Delta \theta^{sep}_{m} > 0$, then there exists angular segments below the $3$ dB levels of the beams and when $\Delta \theta_m^{sep} >> 0$ beam nulls are observed in some directions. The beams can be broadened such that their broadened HPBW points come closer and $\Delta \theta^{sep}_{m}$ is reduced,
in which case all the directions in the FOV would be in the vicinity of $3$ dB beamwidth of the beams as illustrated in Fig. \ref{fig:beams_hpbw}.   
\\

This adaptation has to be done for every sensing frame in real-time and can be accomplished by using parameterized windows as adaptive filters. Slepian $et \; al.$ examined the time-frequency concentration problem and showed that the Slepian window has the property of maximal concentration in both time as well as frequency \cite{SlepianPSWF1} \cite{LandauPSWF2} \cite{SlepianPSWF5}. Subsequently, Kaiser came up with an easier approximation to the above optimal window called the Kaiser window. As we are interested in maximum mainlobe energy concentration in the angular domain along with least number of elements in the elements domain, Kaiser window is chosen to broaden the beams.    
\\

Consider the Kaiser window, with the parameter $\beta$, given by
\begin{equation}\label{eq:kaiser_defn}
w[n] = \frac{I_0 \left(  \beta \sqrt{1 - \left( \frac{n - \frac{N}{2}}{\frac{N}{2}}  \right)^2} \right)}{I_0 \left( \beta \right)}, \; \; \; 0 \leq n \leq N,
\end{equation}
where \( I_0 \) is the zeroth-order modified Bessel function of the first kind and $N$ is the window length in samples.

From an FIR filter design perspective, given the filter specifications: $A_{sl}$ - the required sidelobe attenuation (in dB) and $\Delta\omega$ - the transition bandwidth, a suitable window can be designed with window length $N$ and $\beta$ given by \cite{kaiser1974nonrecursive}
\begin{equation}\label{eq:N_active}
    N = \frac{(A_{sl} - 7.95)}{2.285\Delta\omega}
\end{equation}
and
{\small
\begin{equation} \label{eq:beta_calc}
  \beta =
    \begin{cases}
      0, \;\; A_{sl} < 21\\
      0.5842(A_{sl} - 21)^{0.4} + 0.07886(A_{sl} - 21) 
      ,\;\;21 \leq A_{sl} \leq 50\\
      0.1102 (A_{sl} - 8.7), \;\; A_{sl} > 50 .
    \end{cases}      
\end{equation}
}
\\
Since $A_{sl}$ is not a specification explicitly defined in our context, (\ref{eq:beta_calc}) can be rewritten by substituting for $A_{sl}$ using (\ref{eq:N_active}) as 
{\footnotesize
\begin{equation}\label{eq:beta_N_active_calc}
  \beta =
    \begin{cases}
      0, \; \; N\Delta\omega < 5.7122\\
      0.5842(2.285 N \Delta\omega - 13.05)^{0.4} + 0.07886(2.285 N \Delta\omega - 13.05), 
      \\ \; \; \; \; \; \; \; \;5.7122 \leq N \Delta\omega \leq 18.4026\\
     0.1102 (2.285 N \Delta\omega - 0.75), \; \; N \Delta\omega > 18.4026 \;\;.
    \end{cases}
\end{equation}
}
In the present context, $MN_{stg}$ adjacent filters can be designed in real-time and applied on the beams such that the $m^{th}$ filter's transition bandwidth
\begin{equation}\label{eq:req_fnbw}
\Delta\omega^{m} > 2\Delta\theta^{\text{hpbw}}_{m} + 2\max\left( \frac{\Delta \theta^{sep}_{m-1}}{2}, \frac{\Delta \theta^{sep}_{m}}{2} \right).\end{equation}

The actual values of $\Delta\omega^{m}$ can be application specific and larger the values, flatter the cumulative array factor and vice-versa.
\\

\begin{figure}[htbp]
\centering
\includegraphics[width=0.5\textwidth]{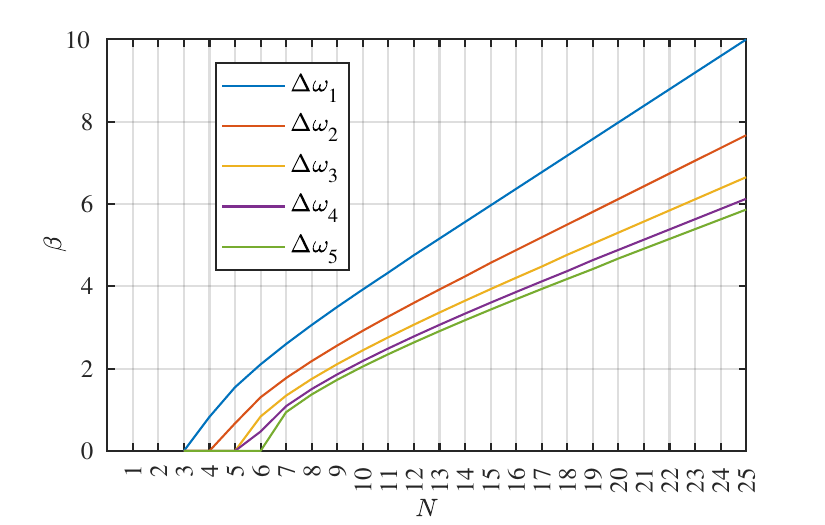}
\caption{\small{\label{fig:n_beta_relation_illustration} Different combinations for $N$, $\beta$ can meet a given FNBW separation $\Delta\omega$. In this plot $\Delta\omega_1 > \Delta\omega_2 > \dots >\Delta\omega_5$. }}
\end{figure}

For the desired mainlobe width ($\Delta\omega$) to be accomplished, it is evident from (\ref{eq:beta_N_active_calc}) that various combinations of $(N, \beta)$ can be employed. This is observed by plotting the values of $\beta$ obtained from (\ref{eq:beta_N_active_calc}) for a given value of $\Delta\omega$ for varying window lengths $N$ in Fig.~\ref{fig:n_beta_relation_illustration}.
We first point out that the values of $\beta$ increase with increasing $\Delta\omega$, to provide the desired broadening of the beamwidth. For a fixed $\Delta\omega$, the values of $\beta$ vary with the chosen window length. Although the number of antennas connected to the analog beamformer is fixed to $L$, it is possible to accomplish beam broadening by reducing the active number of antenna elements to $N\leq L$ along with the appropriate choice of $\beta$. However, reducing the active number of antenna elements to a very low value will reduce the array gain in the beam direction for the transmitted signal. One can therefore choose $0.5 L< N\leq L$ for practical designs.
\\

Denoting $\mathbf{w}_m = [w_m[0], \ldots, w_m[N-1]]$ as the $m^{th}$ filter's corresponding window vector, the windowed beamformer weights for each of the $m = 1 \;\text{to}\; MN_{stg}$ beams can be computed apriori at the onset of the frame as
\begin{equation}\label{eq:windowed_beamformer_weights}
\overline{\mathbf{b}}(\theta_m) =\mathrm{diag}\{\mathbf{w}_m\}\mathbf{b}(\theta_m),
\end{equation}
where $\mathbf{b}(\theta_m)$ is as defined in (\ref{eq:beamformer_weights}).
\subsection{Transmit and Receiver processing}
The $MN_{stg}$ beams are divided into $N_{stg}$ stages such that $M$ beams are transmitted and received in each stage. In the $j^{th}$ stage, the windowed beamforming weights, $\overline{\mathbf{b}}^j(\theta_m)$ for $m = 1 \;\text{to}\; M$ beams, computed apriori in (\ref{eq:windowed_beamformer_weights}), are programmed to the analog beamformer. The signal output of the $m^{th}$ beamformer of the $j^{th}$ stage is given by
\begin{equation}\label{eq:tx_signal_j_m_with_window}
\overline{\mathbf{x}}^j_m(t) = \overline{\mathbf{b}}^j(\theta_m)\mathbf{x}_m^j(t).
\end{equation} 
The signal transmitted by all the $M$ RF chains together in the $j^{th}$ stage is given by
\begin{align} \label{eq:tx_signal_j_with_window}
\mathbf{\overline{x}}^{j}(t) &= [\overline{\mathbf{x}}_1^j(t)^T\ldots, \overline{\mathbf{x}}_m^j(t)^T]^T \nonumber \\
                             &= \mathbf{B}^j_w \mathbf{S}(t), \in \mathbb{C}^{LM \times 1},
\end{align}
where $\mathbf{B}^j_w = \mathrm{diag}\left([\mathbf{\overline{b}^j}(\theta_1)^T,\ldots, \mathbf{\overline{b}^j}(\theta_M)^T]^T\right)$ is the windowed beamspace transmit matrix of the $j^{th}$ stage.

The corresponding windowed array factor for $M$ beams in the $j^{th}$ stage is 
\begin{equation}
   \mathbf{f}_{\mathbf{w}}^{j}(\theta) = \sum_{m=1}^{M} \sum_{k=1}^{L} \mathbf{w}^j_{m}[k] \; e^{-j2\pi\frac{d}{\lambda}k(\sin{\theta} - \sin{\theta_m)}}
\end{equation}
and the overall combined windowed array factor for $N_{stg}$ stages is
\begin{equation}
   \mathbf{f}^\text{overall}_{\mathbf{w}}(\theta) = \sum_{j=1}^{N_{stg}}  \mathbf{f}_{\mathbf{w}}^{j}(\theta) \;\; .
\end{equation}

The RF chains are essentially analog beamformers that give beamformed output and hence are not suitable during signal reception for DoA estimation. We therefore consider a phased array receiver with $L_R$ elements that receives signals from the $N_{stg}$ stages sequentially. 
\\

For the $\mathbf{\overline{x}}^{j}(t)$ transmitted pulses during the $j^{th}$ stage, let $\mathbf{Z}^j\in \mathbb{C}^{L_R\times P\times Q}$ be the radar data cube received for $P$ pulses and $Q$ fast-time samples. Since different beams are active over the $j = 1 \; \text{to}\; N_{stg}$ stages, the received signals from all these stages is added to form the received signal from the entire FOV as
\begin{equation}\label{eq:recv_signal}
\mathbf{Z} = \sum_{j=1}^{N_{stg}}\mathbf{Z}^j
\;\;\; \in \mathbb{C}^{L_R\times P\times Q}.
\end{equation}

Target detection is performed on this data along the fast time to identify the range indices corresponding to the targets. At a specific detected range index $q$, the array data matrix $\mathbf{Z}_q \in \mathbb{C}^{L_R\times P}$ will be used for angle estimation. A wide range of DoA estimation techniques [14-16] can be used for angle estimation. In this study, Capon's estimator is employed to investigate the effect of transmit beam broadening on angle estimation. The corresponding array covariance matrix can be estimated as
\begin{equation}\label{eq:cov_matrix}
\widehat{\mathbf{R}}_{zz} = \frac{1}{PN_{stg}}\mathbf{ZZ}^H.
\end{equation}
and the angular spectrum is given by
\begin{equation}\label{eq:j_stage_capons_spectrum}
\mathbf{P}(\theta) = \frac{1}{\mathbf{a_R}^H(\theta)\widehat{\mathbf{R}}_{zz}^{-1}\mathbf{a_R}(\theta)},
\end{equation}
where $\mathbf{a_R}(\theta)$ is the array steering vector for the received phased array. It is important to note that DoA estimation is performed only once after accumulating data over all the stages.
\\

To reiterate, the proposed multi-stage adaptive windowing technique helps to reliably sense the scene within a sensing frame using two distinct strategies, 
\begin{enumerate}
\item Firstly, with the available $M$ RF chains, the number of beams is increased to a suitable $MN_{stg}$ which allows stage-wise sensing of the FOV within the current sensing frame duration $T_{S_i}~$s. 
\item Second, even with $MN_{stg}$ logical beams, sensing all the directions consistently within the FOV is not assured when the number of RF chains available for sensing - $M$ or number of stages - $N_{stg}$ decreases in some frames. Adaptive windowing identifies the required window parameters for the $MN_{stg}$ beams in real time and broaden them to accomplish consistent performance across the FOV.
\end{enumerate}

\subsection{Application and Practical Considerations}

\begin{algorithm}
{ \small
	\caption{Proposed Multi-stage Adaptive Windowing}
    \label{alg:multi_stage_adaptive_windowing}
	\begin{algorithmic}[1]
        \Require $T_{CPI}$, FOV = $[\theta_{start}$, $\theta_{end}]$
        \For {every frame} \Comment{Opt1}
        \State Get $T_{S_i}$ and $M = R_{s_i}$ from the scheduler
		\State Calculate $N_{stg}$ (using (\ref{eq:N_stg}))
        \State \parbox[t]{\dimexpr\linewidth-\algorithmicindent}{%
        Construct $MN_{stg}$ beams at positions $\theta_m$ given by (\ref{eq:mnstgth_beam_position}) 
        with beamformer weights $b(\theta_m)$ given by (\ref{eq:beamformer_weights}) 
        }
        \State
        \For {$m = 1$ to $MN_{stg}$} 
        \State \parbox[t]{\dimexpr\linewidth-\algorithmicindent}{%
        Compute HPBWs $\Delta\theta^{\text{hpbw}}_{m}$ and angular separations $\Delta\theta^{\text{sep}}_{m}$ (\ref{eqn:hpbw_beamwidth_approx}) and (\ref{eqn:angular_sep_beams})
        }
        \State
        \If {$\Delta\theta^{\text{sep}}_{m} > 0$}
        \State Compute required main lobe width $\Delta\omega^{m}$ (\ref{eq:req_fnbw})
        \State For a suitable $N_m$ compute $\beta_m$ (\ref{eq:beta_N_active_calc})
        \State Compute and store $\overline{\mathbf{b}}(\theta_m)$ (\ref{eq:windowed_beamformer_weights})
        \Else 
        \State no windowing needed. Use $\mathbf{b}(\theta)$.
        \EndIf
        \State
        \EndFor \Comment{Opt2}
        \State
        \For {$j = 1$ to $N_{stg}$}
        \State Transmit $P$ pulses of $\mathbf{\overline{x}}^{j}(t)$ (\ref{eq:tx_signal_j_with_window})  
        \State Receive responses $\mathbf{Z}^j$ and add (\ref{eq:recv_signal}) 
        \EndFor
        \State
        \State \parbox[t]{\dimexpr\linewidth-\algorithmicindent}{%
        Detection algorithm gives range indices $q$. Extract $\mathbf{Z}_q$ from the data cube.
        }
        \State Estimate the covariance matrix  $\widehat{\mathbf{R}}_{zz}$ (\ref{eq:cov_matrix})
        \State Estimate the angular spectrum $\mathbf{P}(\theta)$ (\ref{eq:j_stage_capons_spectrum})
        \State Find peaks in $\mathbf{P}(\theta)$ to find target directions
        \EndFor
	\end{algorithmic} 
    }
\end{algorithm}

Algorithm \ref{alg:multi_stage_adaptive_windowing} provides the core implementation outline of the proposed multi-stage adaptive windowing technique. A few implicit design ideas, silicon-based mmWave phased array IC features and system optimization techniques relevant to the algorithm are briefly mentioned which might aid system designers while designing the overall DFRC application.
\begin{enumerate}

\item The DFRC application should, at some system level, have separate but concurrently running communication and sensing functionalities, eg. threads, dedicated processing units etc. and suitable low latency message/data sharing mechanisms between communication and sensing functionalities.

\item From label Opt1 to label Opt2 in the algorithm, the implementation can be optimized using beamforming ICs with integrated beam memory arrays. Since the scheduler granularity is defined by an appropriate slot duration, $T_{S_i}$ can be viewed as some integral number of slots and $T_{CPI}$ is a constant. $M$ can change between $1$ and $R_{T}$ which gives a pre-defined range of $MN_{stg}$ values. The associated angles, beamforming weights and windowed beamforming weights can be pre-computed and stored in beam memory arrays thus reducing the labelled operations to simple look-ups.

\item The latest RFIC chips permit programming both the amplitude and phase of the beamforming weights. This enables easy application of the window during transmit beamforming.

\item Finer optimizations are possible, e.g., by adjusting number of active elements, $N_m$, to an appropriate value less than $L$, $T_{CPI}$ can be brought down at the cost of array gain.

\end{enumerate}

\section{Simulation Studies}

For the subsequent simulation studies, we consider a mmWave DFRC BS with pulsed wave radar and parameters as below
{\small
\begin{center}
\begin{tabular}{ | c | c |}
\hline
Total Available RF Chains - $R_T$ & $25$ \\
\hline
Antennas per RF Chain - $L$   & $25$ \\
\hline
Operating Frequency - $f0$  & $28$ GHz \\
\hline
Intermediate Frequency - $f_{if}$  & $20$ MHz \\
\hline
Sampling Frequency - $f_{s}$  & $150$ MHz \\
\hline
Bandwidth - $B$   & $20$ MHz \\
\hline
Pulse Duration - $T_p$ & $1$ $\mu$s \\
\hline
Pulse Repetition Interval - $T_{pri}$ & $20$ $\mu$s \\
\hline
Pulses per CPI - $P$   & $32$ \\
\hline
Antennas in Receiver - $L_R$   & $8$ \\
\hline
\end{tabular}
\end{center}
}
The received signal power is modelled using the Radar range equation and the thermal noise at the receiver is modelled separately as $kTB$ where $k$ is the Boltzmann constant, $T = 300$ K is the temperature and $B$ is the bandwidth.
\subsection*{Study 1: Cumulative beam pattern and angular spectrum}

\begin{figure}[htbp]
\centering
\includegraphics[width=0.5\textwidth]{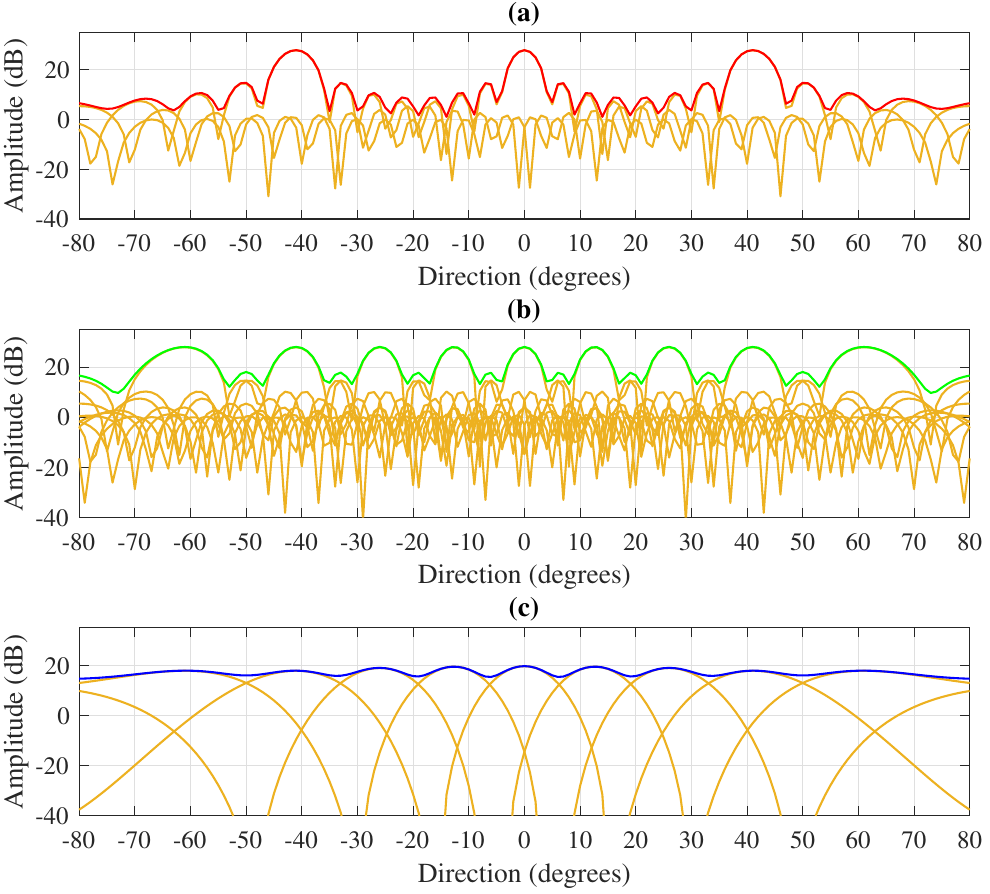}
\caption{\small{\label{fig:25s_3b_multistage_win_beampatterns} Individual beam patterns (orange) and corresponding cumulative beam patterns for (a) $M=3$ non-windowed beams, (b) $MN_{stg}=9$ non-windowed beams and (c) $MN_{stg}=9$ windowed beams.}}
\end{figure}

We illustrate the technique when the number of RF chains available for sensing is $M = 3$ and the number of stages $N_{stg} = 3$. We present the individual and cumulative beam patterns for $M=3$ beams, $MN_{stg}=9$ non-windowed beams, and $MN_{stg}=9$ windowed beams in Fig. \ref{fig:25s_3b_multistage_win_beampatterns}.

In Fig. \ref{fig:25s_3b_multistage_win_beampatterns} (a), while naively sensing with $3$ beams, we note that a number of beam null positions are present in the FOV where the power drops by nearly $30$ dB. While this is remediated to a certain extent in Fig. \ref{fig:25s_3b_multistage_win_beampatterns} (b) by sensing stage-wise and increasing the beam count to $9$, the cumulative beam pattern still shows power drop in certain directions which results in inconsistent sensing. In Fig. \ref{fig:25s_3b_multistage_win_beampatterns} (c), the $9$ beams are windowed adaptively such that the FOV coverage is more uniform, with a trade-off in the power along the beam peak directions compared to the non-windowed case.

\begin{figure}[htbp]
\centering
\includegraphics[width=0.5\textwidth]{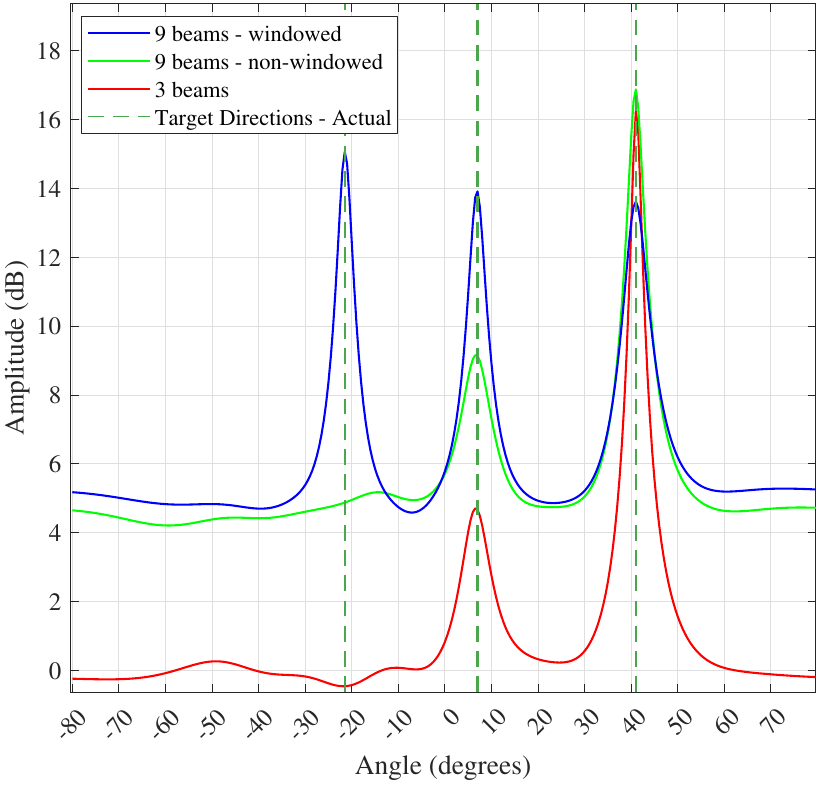}
\caption{\small{\label{fig:25s_3b_3stg_3tgt_illustration} Angular spectra for $3$ targets placed at positions $-21.5^{\circ}$, $+7.0^{\circ}$ and $+41.0^{\circ}$. } }
\end{figure}
We consider three targets at comparable ranges of $450$ m, $500$ m and $550$ m, velocities of $1$ m/s, $-2$ m/s and $3$ m/s and identical radar cross-section (RCS) of $2$ dBsm. The $3$ targets are positioned at $+41^{\circ}$, $+7^{\circ}$, and $-21.5^{\circ}$. These directions are chosen such that they are positioned along the beam peak, off-beam peak and beam null, respectively. For the three transmit beam configurations discussed earlier, DoA estimation is then performed on the received signal. The resulting angular spectra for the three beam patterns are shown in Fig. \ref{fig:25s_3b_3stg_3tgt_illustration}. In both $M$ beams and $MN_{stg}$ cases, the targets at beam peak and average beam power directions have peaks distinctly above the noise floor and are estimated correctly. However, the peak for the target at the beam null direction $-21.5^{\circ}$ is buried in the noise floor in the above two configurations, whereas it is estimated correctly in the $MN_{stg}$ windowed case due to the uniform performance across the FOV.

\subsection*{Study 2: Beam patterns for different $MN_{stg}$ and window adaptation}

\begin{figure}[htbp]
\centering
\includegraphics[width=0.5\textwidth]{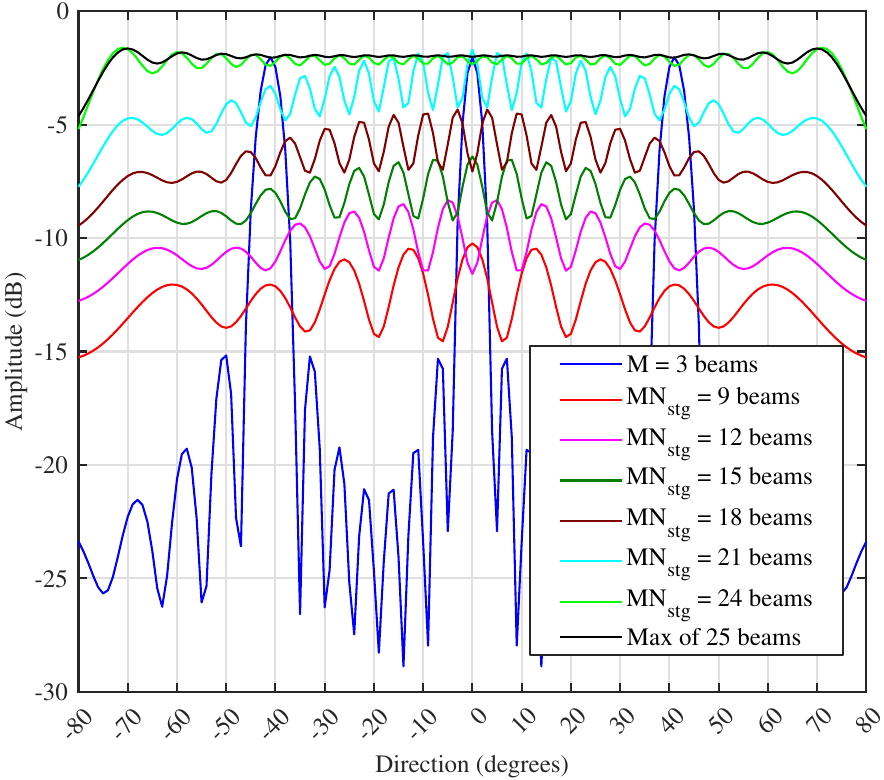}
\caption{\small{\label{fig:beams_mn_stages_convergence} Beam patterns for $M = 3$ beams, $M = 25$ beams and various $MN_{stg}$ windowed beams from 9 to 24.  As $MN_{stg}$ increases, the beam pattern approaches that of $M = 25$ beams. }}
\end{figure}

In this study, we examine the effect of $MN_{stg}$ on the cumulative windowed beam pattern and window parameters. We again consider number of stages $N_{stg} = 3$ and RF chain with $L = 25$ antennas. The available RF chains $M$ are varied from $3$ to $8$ to provide $MN_{stg} = 9, 12, 15, 18, 21$ and $24$ beams, respectively. The cumulative windowed beam patterns thus obtained are compared with the stage-less, raw beam pattern when $M = 3$ beams and the optimal beam pattern when $M = 25$ beams. The results are shown in Fig. \ref{fig:beams_mn_stages_convergence}.

As the number of available RF chains $M$ is increased, $MN_{stg}$ also increases. Thus, there is an increased overall gain across the FOV with this increase in the logical number of beams. Additionally, adaptive windowing results in a lower power variation in the beam patterns across the FOV thus avoiding deep nulls which can be seen in the lower limiting case of $M = 3$ beams.  As the number of available RF chains are increased, the multi-stage windowed beam pattern approaches that of the optimal beam pattern with $M = 25$ beams. 
\\

\begin{figure}[htbp]
\centering
\includegraphics[width=0.5\textwidth]{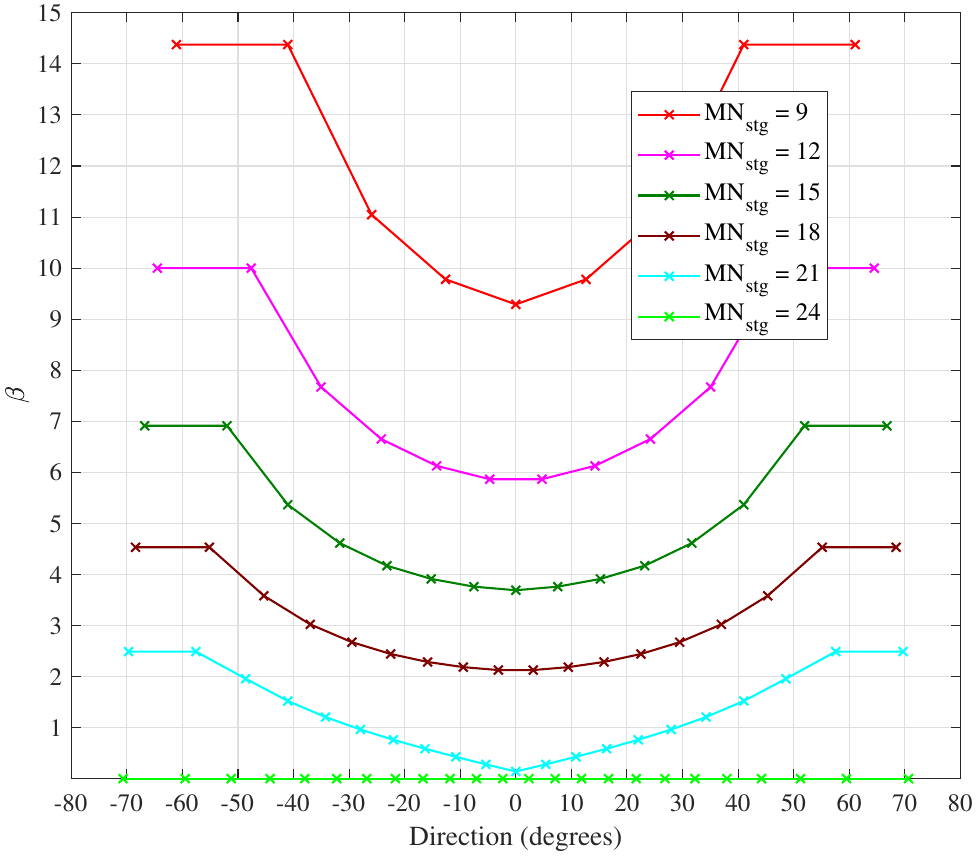}
\caption{\small{\label{fig:beta_values_convergence} Beta values of Kaiser windows applied during adaptive windowing for various $MN_{stg}$ beams from 9 to 24. The applied window approaches a rectangular window when $MN_{stg}$ is high. }}
\end{figure}

For each case of $MN_{stg}$ studied, the window has to vary from beam to beam in accordance with its position to maintain consistent performance across the FOV. The window parameters assigned for all the $MN_{stg}$ beams are plotted in Fig. \ref{fig:beta_values_convergence}.

When $MN_{stg}$ is less, the separation between the beams is more as the beams are further apart. Hence, higher $\beta$ values are necessary to reduce the HPBW separation.  With increasing RF chains $M$, progressively higher number of beams are available and their separation in the angular domain decreases. Thus, smaller values for $\beta$ are applied. It can also be noted from  Fig. \ref{fig:beta_values_convergence} that different $\beta$ values are computed for different beams. This is due to the increase in beam widths and beam separation when beamforming away from the broadside of the array. Importantly, when the number of available RF chains is sufficiently high ($M \geq 8$, in this case) the Kaiser window approaches a rectangular window and the technique becomes equivalent to a full FOV sequential scan.
\subsection*{Study 3: FOV Coverage - RMSE and Probability of Detection}

In the first part of this study, we perform a controlled experiment to study the probability of detection (PD) and the root mean-squared error (RMSE) of the target DoA estimates over the entire FOV. 

\begin{figure}[htbp]
\centering
\includegraphics[width=0.5\textwidth]{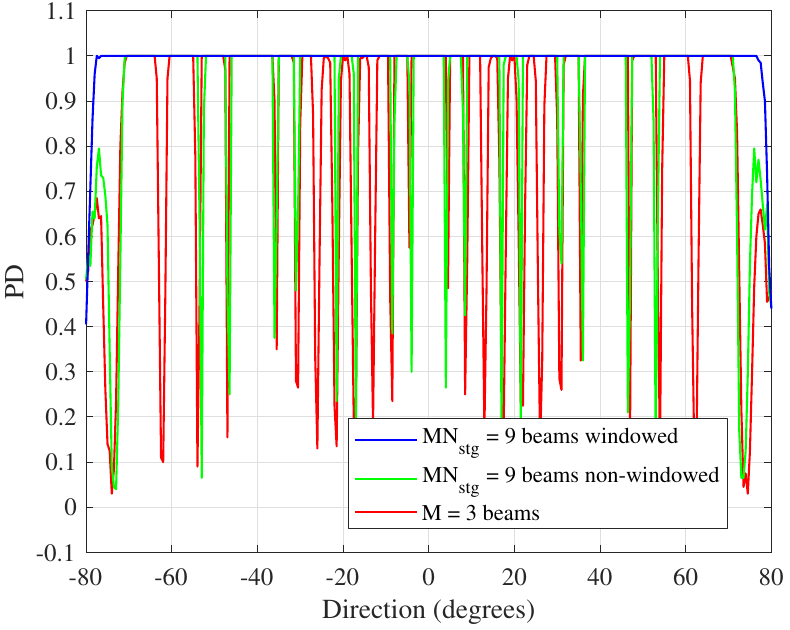}
\caption{\small{\label{fig:PD_Directions} Probability of Detection across the FOV for the three configurations. }}
\end{figure}

\begin{figure}[htbp]
\centering
\includegraphics[width=0.5\textwidth]{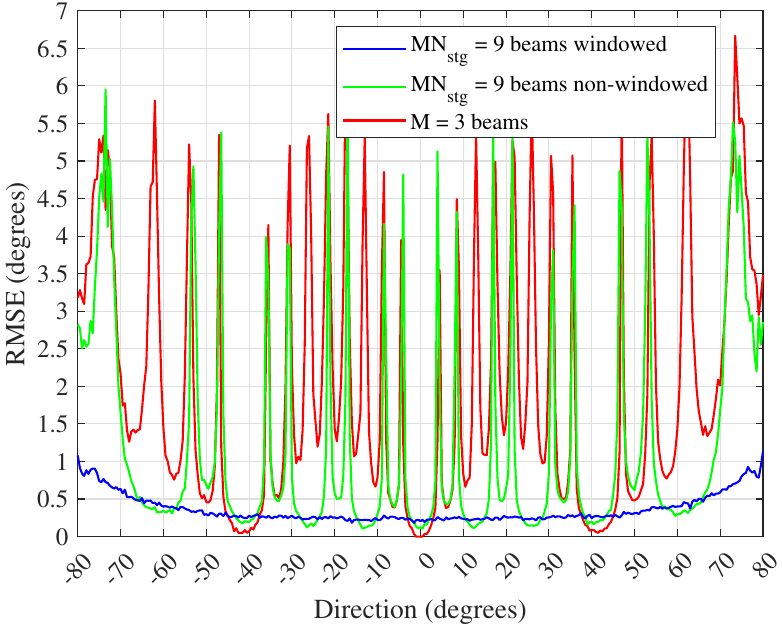}
\caption{\small{\label{fig:RMSE_Directions} RMSE across the FOV for the three configurations. }}
\end{figure}

We consider a single target with a fixed range, velocity and RCS of $250$ m, $1$ m/s and $2$ dBsm respectively, and vary its position within the FOV of $[-80^{\circ}, +80^{\circ}]$ in steps of $0.5^{\circ}$. The signal-to-noise ratio (SNR) of the received signal is maintained at 10 dB, and for each position, a Monte-Carlo simulation is run over 200 iterations and DoA estimation of this target is performed using the earlier three configurations, viz.,  $M = 3$ beams, $M = 3$ beams and $N_{stg} = 3$ stages and  $M = 3$ beams and $N_{stg} = 3$ stages with windowing. If the target direction estimate differs by more than $10^{\circ}$, we declare the target as not detected. Else, we consider the estimate for computing the RMSE. The PD and RMSE for the three configurations are plotted in Figures \ref{fig:PD_Directions} and \ref{fig:RMSE_Directions}, respectively.
\\

For the configuration with $M = 3$ beams, the PD is low in the angular sectors where beam nulls are present. Correspondingly, RMSE is high in these regions. Although using $MN_{stg} = 9$ beams shows improved PD and RMSE, performance is inconsistent in between the beams due to inconsistent SNR in those directions. When windowing is employed on $MN_{stg}$ beams in the third configuration, consistent target detection over the entire FOV is observed in Fig. \ref{fig:PD_Directions} due to a more consistent beam pattern through out. Consequently, RMSE is alike and lower than the other two configurations in most directions as seen in Fig. \ref{fig:RMSE_Directions}.
\\

In the second part of the study, we evaluate the technique on randomized target parameters and different SNRs.
\begin{figure}[htbp]
\centering
\includegraphics[width=0.5\textwidth]{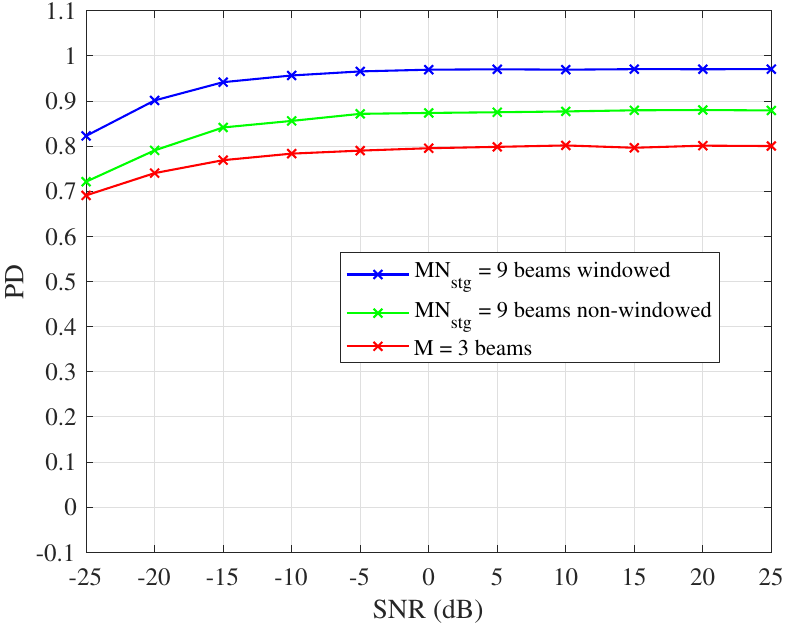}
\caption{\small{\label{fig:snr_avg_pd} Probability of Detection for the three configurations for SNRs ranging from $-25$ dB to $25$ dB. }}
\end{figure}

\begin{figure}[htbp]
\centering
\includegraphics[width=0.5\textwidth]{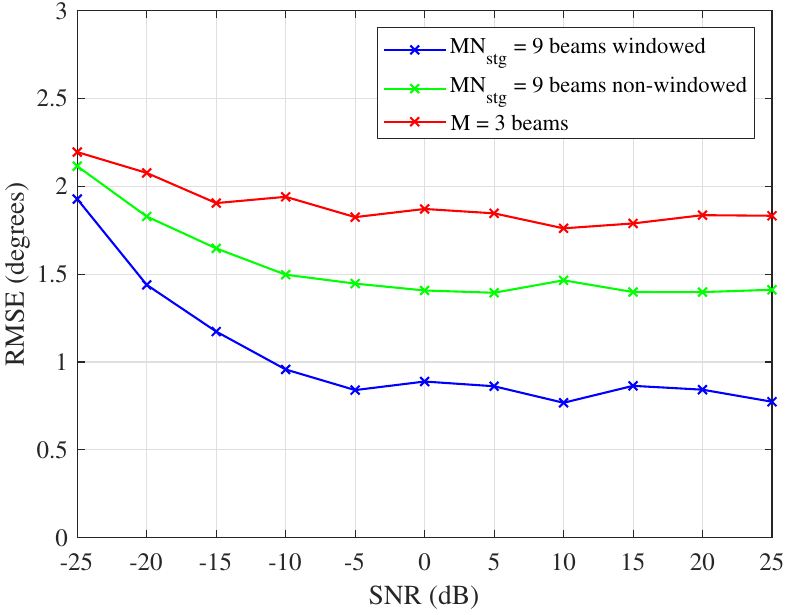}
\caption{\small{\label{fig:snr_avg_rmse} RMSE for the three configurations for SNRs ranging from $-25$ dB to $25$ dB. }}
\end{figure}

We consider three targets drawn from a uniform distribution with range, velocity and direction between $250$m to $1500$m, $-30$m/s to $+30$m/s and $-80^{\circ}$ to $+80^{\circ}$ respectively, to construct a scenario. For each scenario, the SNR of the received signal is varied from $-25$ dB to $25$ dB with the receiver noise separately modelled as $kTB$. Monte-Carlo simulation is run over 2000 such scenarios and target DoAs are estimated using $M = 3$ beams, $MN_{stg} = 9$ beams and $MN_{stg} = 9$ beams with windowing. In order to avoid the target range detection algorithms affecting our study, we assume the knowledge of the range bin corresponding to the targets before DoA estimation. The PD and RMSE of the DoA estimates are studied for the three configurations as shown in Figures \ref{fig:snr_avg_pd} and \ref{fig:snr_avg_rmse} respectively.

Figure \ref{fig:snr_avg_pd} shows the target DoA detection performance for the three configurations. The PD for $M=3$ beams and $MN_{stg}=9$ without windowing does not reach the performance of $MN_{stg}=9$ with windowing even at higher SNRs. This is because some targets amongst all the trials are positioned at, or close, to beam null directions. On the contrary, the proposed technique provides reliable target DoA detections both in the beam null as well as other directions within the FOV and therefore, the PD for $MN_{stg}=9$ with windowing is close to $1$. 
\\

The corresponding RMSE for the three configurations are plotted in Fig. \ref{fig:snr_avg_rmse}. Again, the proposed technique exhibits superior performance over the other two configurations because of the increased signal power in trials where targets are positioned in the beam null directions.  

\subsection*{Study 4: Dynamic RF Chain Allocation - RMSE and Probability of Detection}

\begin{figure}[htbp]
\centering
\includegraphics[width=0.5\textwidth]{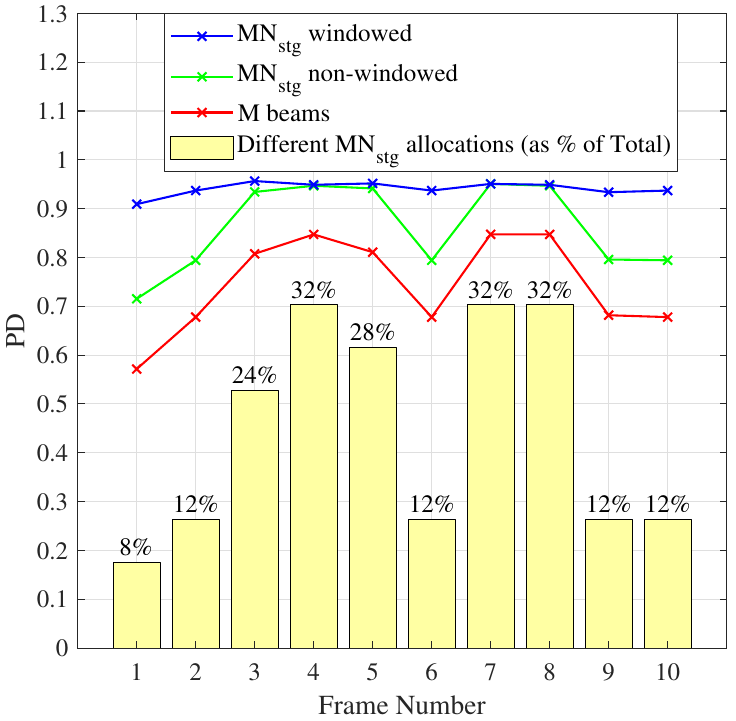}
\caption{\small{\label{fig:10_frames_poisson_avg_pd} Frame-wise PD for varying RF chain allocations per frame. }}
\end{figure}

\begin{figure}[htbp]
\centering
\includegraphics[width=0.5\textwidth]{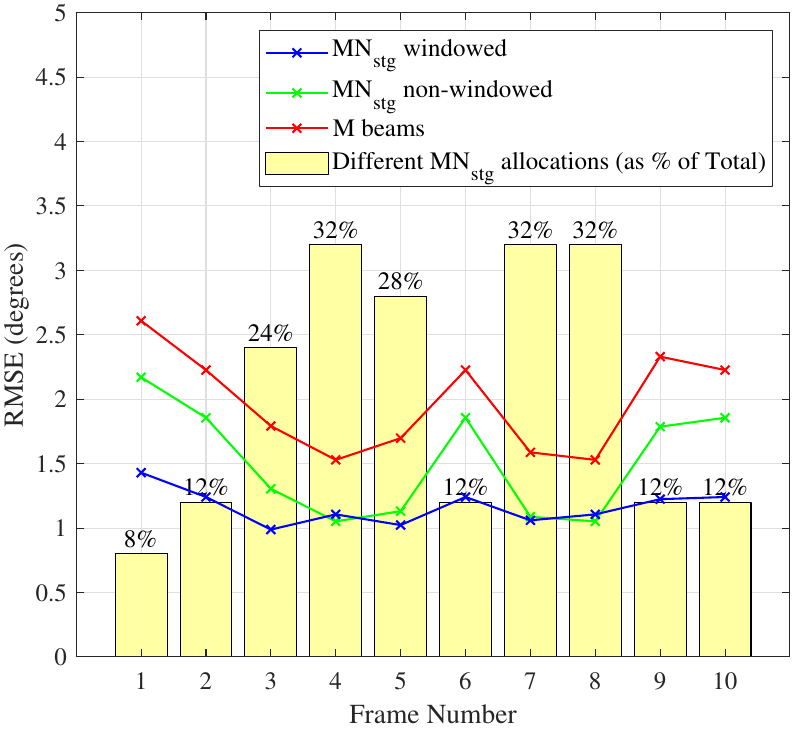}
\caption{\small{\label{fig:10_frames_poisson_avg_rmse} Frame-wise RMSE for varying RF chain allocations per frame. }}
\end{figure}

Finally, we simulate the real-time dynamism in RF chain allocation for sensing and evaluate the performance of the technique in this dynamic scenario. The RF chain allocation per frame is modeled as a Poisson process with $\lambda = 5$. This translates to $5$ out of $25$ RF chains (i.e., $20\%$) allocated for sensing per frame on an average. An observation of this process is picked for a length of 10 frames. A fixed sensing frame duration ($T_{S_i}$s) is considered in this study such that number of stages $N_{stg} = 3$. The received signal SNR is maintained at $10$ dB. For each observation, we consider three targets drawn from a uniform distribution with range, velocity and direction between $250$m to $1500$m, $-30$m/s to $+30$m/s and $-80^{\circ}$ to $+80^{\circ}$ respectively, to construct a scenario. Monte-Carlo simulation is run over 2000 such scenarios and DoAs are estimated using the above three techniques. The average PD values and RMSE of the DoA estimates along with the allocated number of RF chains for each frame are shown in Figures \ref{fig:10_frames_poisson_avg_pd} and \ref{fig:10_frames_poisson_avg_rmse} respectively. 
\\

It is observed that for the configuration with $M$ beams, the PD drops when the number of RF chains allocated for sensing reduces in a given frame. Correspondingly, RMSE is high in these regions. Although using $MN_{stg}$ beams shows improved PD and RMSE, the performance is still inconsistent when the number of RF chains allocated per frame varies. For $MN_{stg}$ beams with windowing, PD is consistent with higher probability values across frames despite the number of RF chains varying as observed in Fig. \ref{fig:10_frames_poisson_avg_pd}. Further, the RMSE remains consistently low over all the frames as seen in Fig. \ref{fig:10_frames_poisson_avg_rmse}. It is therefore demonstrated that the proposed technique provides a mechanism to adapt to varying resources for consistent sensing and DoA performance over the entire FOV.

\section{Conclusion}
In this paper, a conceptual DFRC scheduler which shares RF chains between radar and communication functions is considered. In such a scheduler, if communication is prioritized over sensing, the RF chains and the time slots allocated to sensing are less and varying. We proposed a multi-stage, adaptive windowing technique for transmit analog beamformer RF chains that works in the above settings to sense the entire FOV and estimate DoA by adapting to varying resources and time. DoA estimation consistency and dynamic adaptability of the proposed technique over existing strategies was demonstrated through extensive numerical simulations and measurements including RMSE and probability of detection.

\nocite{Jeffrey2009phased}
\nocite{Van2002optimum}
\nocite{Fuhrmann2008transmit}
\nocite{Stoica2007probing}
\nocite{Liu2020joint}
\nocite{Abigael2024Angular}
\nocite{Kim2013multi}
\nocite{richards2014fundamentals}
\nocite{harris1978windows}
\nocite{prabhu2014window}
\nocite{kaiser1974nonrecursive}
\nocite{Elbir2023BeamformingAdvances}

\bibliographystyle{IEEEtran}
\bibliography{references.bib}

\begin{thebibliography}{10}
\providecommand{\url}[1]{#1}
\csname url@samestyle\endcsname
\providecommand{\newblock}{\relax}
\providecommand{\bibinfo}[2]{#2}
\providecommand{\BIBentrySTDinterwordspacing}{\spaceskip=0pt\relax}
\providecommand{\BIBentryALTinterwordstretchfactor}{4}
\providecommand{\BIBentryALTinterwordspacing}{\spaceskip=\fontdimen2\font plus
\BIBentryALTinterwordstretchfactor\fontdimen3\font minus
  \fontdimen4\font\relax}
\providecommand{\BIBforeignlanguage}[2]{{%
\expandafter\ifx\csname l@#1\endcsname\relax
\typeout{** WARNING: IEEEtran.bst: No hyphenation pattern has been}%
\typeout{** loaded for the language `#1'. Using the pattern for}%
\typeout{** the default language instead.}%
\else
\language=\csname l@#1\endcsname
\fi
#2}}
\providecommand{\BIBdecl}{\relax}
\BIBdecl

\bibitem{zheng2019dfrc}
L.~Zheng, M.~Lops, Y.~C. Eldar, and X.~Wang, ``Radar and communication
  coexistence: An overview: A review of recent methods,'' \emph{IEEE Signal
  Processing Magazine}, vol.~36, no.~5, pp. 85--99, 2019.

\bibitem{Hassanien2019DualFunctionRC}
A.~Hassanien, M.~G. Amin, E.~Aboutanios, and B.~Himed, ``Dual-function radar
  communication systems: A solution to the spectrum congestion problem,''
  \emph{IEEE Signal Processing Magazine}, vol.~36, pp. 115--126, 2019.

\bibitem{Liu2020dfrc}
F.~Liu, C.~Masouros, A.~P. Petropulu, H.~Griffiths, and L.~Hanzo, ``Joint radar
  and communication design: Applications, state-of-the-art, and the road
  ahead,'' \emph{IEEE Transactions on Communications}, vol.~68, no.~6, pp.
  3834--3862, 2020.

\bibitem{Liu2017TowardDR}
F.~Liu, L.~Zhou, C.~Masouros, A.~Li, W.~Luo, and A.~P. Petropulu, ``Toward
  dual-functional radar-communication systems: Optimal waveform design,''
  \emph{IEEE Transactions on Signal Processing}, vol.~66, pp. 4264--4279, 2017.

\bibitem{Sturm2011waveform}
C.~Sturm and W.~Wiesbeck, ``Waveform design and signal processing aspects for
  fusion of wireless communications and radar sensing,'' \emph{Proceedings of
  the IEEE}, vol.~99, no.~7, pp. 1236--1259, 2011.

\bibitem{Sadhu2017}
B.~Sadhu, Y.~Tousi, J.~Hallin, S.~Sahl, S.~K. Reynolds, {\"O}.~Renstr{\"o}m,
  K.~Sj{\"o}gren, O.~Haapalahti, N.~Mazor, B.~Bokinge, G.~Weibull,
  H.~Bengtsson, A.~Carlinger, E.~Westesson, J.-E. Thillberg, L.~Rexberg,
  M.~Yeck, X.~Gu, M.~Ferriss, D.~Liu, D.~Friedman, and A.~Valdes-Garcia, ``A
  28-ghz 32-element trx phased-array ic with concurrent dual-polarized
  operation and orthogonal phase and gain control for 5g communications,''
  \emph{IEEE Journal of Solid-State Circuits}, vol.~52, no.~12, pp. 3373--3391,
  2017.

\bibitem{belfiori2012random}
F.~Belfiori, W.~van Rossum, and P.~Hoogeboom, ``Random transmission scheme
  approach for a fmcw tdma coherent mimo radar,'' in \emph{2012 IEEE Radar
  Conference}.\hskip 1em plus 0.5em minus 0.4em\relax IEEE, 2012, pp.
  0178--0183.

\bibitem{Billam1997Time}
E.~Billam, ``The problem of time in phased array radar,'' in \emph{Radar 97
  (Conf. Publ. No. 449)}, 1997, pp. 563--575.

\bibitem{GrahamSubArray2013}
T.~D. Graham and D.~R. Culkin, ``Radar architecture using mimo transmit
  subarrays,'' in \emph{2013 IEEE International Symposium on Phased Array
  Systems and Technology}, 2013, pp. 408--415.

\bibitem{RajSubArrays2012}
S.~Rajagopal, ``Beam broadening for phased antenna arrays using multi-beam
  subarrays,'' in \emph{2012 IEEE International Conference on Communications
  (ICC)}, 2012, pp. 3637--3642.

\bibitem{gershman1999robust}
A.~B. Gershman, ``Robust adaptive beamforming in sensor arrays,'' \emph{Int. J.
  Electron. Commun.}, vol.~53, pp. 305--314, 1999.

\bibitem{Nai2010BeamPatternSynthesis}
S.~E. Nai, W.~Ser, Z.~L. Yu, and H.~Chen, ``Beampattern synthesis for linear
  and planar arrays with antenna selection by convex optimization,'' \emph{IEEE
  Transactions on Antennas and Propagation}, vol.~58, no.~12, pp. 3923--3930,
  2010.

\bibitem{Van2002optimum}
H.~L. Van~Trees, \emph{Optimum array processing: Part IV of detection,
  estimation, and modulation theory}.\hskip 1em plus 0.5em minus 0.4em\relax
  John Wiley \& Sons, 2002.

\bibitem{SlepianPSWF1}
D.~Slepian and H.~O. Pollak, ``Prolate spheroidal wave functions, fourier
  analysis and uncertainty — i,'' \emph{The Bell System Technical Journal},
  vol.~40, no.~1, pp. 43--63, 1961.

\bibitem{LandauPSWF2}
H.~J. Landau and H.~O. Pollak, ``Prolate spheroidal wave functions, fourier
  analysis and uncertainty — ii,'' \emph{The Bell System Technical Journal},
  vol.~40, no.~1, pp. 65--84, 1961.

\bibitem{SlepianPSWF5}
D.~Slepian, ``Prolate spheroidal wave functions, fourier analysis, and
  uncertainty — v: the discrete case,'' \emph{The Bell System Technical
  Journal}, vol.~57, no.~5, pp. 1371--1430, 1978.

\bibitem{kaiser1974nonrecursive}
J.~F. Kaiser, ``Nonrecursive digital filter design using window function,'' in
  \emph{Proceedings of the 1974 IEEE International Symposium on Circuits and
  Systems}.\hskip 1em plus 0.5em minus 0.4em\relax IEEE, 1974, pp. 20--23.

\bibitem{Jeffrey2009phased}
T.~Jeffrey and T.~Jeffrey, \emph{Phased-Array Radar Design: Application of
  Radar Fundamentals}, ser. Electromagnetics and Radar.\hskip 1em plus 0.5em
  minus 0.4em\relax Institution of Engineering and Technology, 2009.

\bibitem{Fuhrmann2008transmit}
D.~R. Fuhrmann and G.~San~Antonio, ``Transmit beamforming for mimo radar
  systems using signal cross-correlation,'' \emph{IEEE Transactions on
  Aerospace and Electronic Systems}, vol.~44, no.~1, pp. 171--186, 2008.

\bibitem{Stoica2007probing}
P.~Stoica, J.~Li, and Y.~Xie, ``On probing signal design for mimo radar,''
  \emph{IEEE Trans. Signal Processing}, vol.~55, no.~8, pp. 4151--4161, 2007.

\bibitem{Liu2020joint}
X.~Liu, T.~Huang, N.~Shlezinger, Y.~Liu, J.~Zhou, and Y.~C. Eldar, ``Joint
  transmit beamforming for multiuser mimo communications and mimo radar,''
  \emph{IEEE Transactions on Signal Processing}, vol.~68, pp. 3929--3944, 2020.

\bibitem{Abigael2024Angular}
A.~Taylor and O.~Rabaste, ``Let it guwo: Waveform optimization for angular
  blanking and robustification in a mimo dual-functional radar-communication
  system,'' \emph{IEEE Transactions on Radar Systems}, vol.~2, pp. 380--390,
  2024.

\bibitem{Kim2013multi}
C.~Kim, T.~Kim, and J.-Y. Seol, ``Multi-beam transmission diversity with hybrid
  beamforming for mimo-ofdm systems,'' in \emph{2013 IEEE Globecom Workshops
  (GC Wkshps)}.\hskip 1em plus 0.5em minus 0.4em\relax IEEE, 2013, pp. 61--65.

\bibitem{richards2014fundamentals}
M.~A. Richards, \emph{Fundamentals of radar signal processing}.\hskip 1em plus
  0.5em minus 0.4em\relax McGraw-Hill Education, 2014.

\bibitem{harris1978windows}
F.~Harris, ``On the use of windows for harmonic analysis with the discrete
  fourier transform,'' \emph{Proceedings of the IEEE}, vol.~66, no.~1, pp.
  51--83, 1978.

\bibitem{prabhu2014window}
K.~M.~M. Prabhu, \emph{Window functions and their applications in signal
  processing}.\hskip 1em plus 0.5em minus 0.4em\relax Taylor \& Francis, 2014.

\bibitem{Elbir2023BeamformingAdvances}
A.~M. Elbir, K.~V. Mishra, S.~A. Vorobyov, and R.~W. Heath, ``Twenty-five years
  of advances in beamforming: From convex and nonconvex optimization to
  learning techniques,'' \emph{IEEE Signal Processing Magazine}, vol.~40,
  no.~4, pp. 118--131, 2023.

\end{thebibliography}

\end{document}